\documentclass[prl,twocolumn,showpacs,preprintnumbers,amsmath,floatfix]{revtex4}
\usepackage{graphicx}
\usepackage{dcolumn}
\usepackage{subfigure}
\usepackage{wrapfig}
\usepackage{cancel}
\usepackage{color}

\begin{document}



\title{ \bf  Effect of dopant atoms on local superexchange in cuprate superconductors:
A perturbative treatment}
\author{ Kateryna Foyevtsova$^1$, Roser Valent\'{\i}$^1$ and P. J. Hirschfeld$^2$}
\affiliation{$^1$Institut f\"ur Theoretische Physik,
Goethe-Universit\"at Frankfurt,
 60438 Frankfurt am Main, Germany \\
$^2$University of Florida, Gainesville, Florida 32611, USA}
\date{\today}

\begin{abstract}
Recent scanning tunneling spectroscopy experiments have
provided evidence that dopant impurities in high-$T_c$
superconductors can strongly modify the electronic structure of
the CuO$_2$ planes nearby, and possibly influence the pairing.  To
investigate this connection,
we calculate the local magnetic superexchange $J$ between
 Cu ions in the presence of dopants within the framework of the three-band Hubbard model, up to fifth order in
perturbation theory.  We
  demonstrate that
the  sign of the change in $J$
 depends on
the relative dopant-induced spatial variation of the atomic levels
in the CuO$_2$ plane, contrary to  results obtained within the
one-band Hubbard model. We discuss some realistic cases and
their
relevance for theories of the pairing mechanism in the cuprates.

\end{abstract}

\maketitle

\section{i. Introduction}

Scanning tunnelling spectroscopy (STS)
experiments on surfaces of several high-$T_c$
materials\cite{Fischer} have discovered a host of fascinating
phenomena, including checkerboard local density of states (LDOS)
modulations and inhomogeneous superconductivity with enormous gap
modulations taking place at the
nanoscale\cite{davisinhom1,davisinhom2,Kapitulnik1,YKohsaka:2007}.
Recently, the size of the local gap was found to be positively
correlated with simultaneously imaged atomic scale defects,
thought to be interstitial oxygen dopants\cite{KMcElroy_05}. This
is a surprising result, since it had been expected that an oxygen,
which donates two holes to the CuO$_2$ plane, would overdope the
system and lead to a smaller gap nearby. The positive correlation
between the positions of the oxygens and the gap led Nunner {\it et
al.}\cite{Nunner_05} to suggest that the dopants might be
increasing the pair interaction locally. This could occur if the
local electronic structure were altered significantly, so as to
modify a spin fluctuation exchange effective interaction, or
possibly a local electron-phonon coupling constant. For example,
it has been observed that a strong correlation exists between the
distance of the apical oxygen from the CuO$_2$ plane in high-$T_c$
materials and the critical temperature\cite{Ohta,Pavarini}, and it
might be imagined that a modulation of this displacement by dopant
atoms could change the pairing interaction locally.

Nunner {\it et al.}\cite{Nunner_05} did not assume any specific
microscopic model, but pointed out simply that the general
assumption of dopants modulating the pair interaction could
explain  a remarkable number of experimental results and
correlations. Within a generalized inhomogeneous Bardeen-Cooper-Schrieffer (BCS) pairing model
adopted in this work, it correctly reproduces the anticorrelation
of coherence peak height and position, the correlation of dopant
position with gap size, and the detailed frequency dependence of
the O:LDOS($\omega$) correlation.   The theory is still
controversial; in almost all treatments of disorder in
superconductors, impurities are assumed to simply scatter
electrons as a screened Coulomb potential, rather than modulate
the pair interaction. There are, however, well-known
exceptions\cite{Earlytau1}, and it is certainly reasonable to
expect modulation of the pair interactions to be largest in
systems like the cuprates where the coherence length is small.
Strong correlations have not been included systematically in the
theory, although some first steps have been made when J. X. Zhu
showed explicitly using an inhomogeneous slave-boson approach that
the proposal of Nunner {\it et al.} that impurities might modulate the
local exchange $J$ was consistent with the STM
observations.\cite{JXZhu:2006}.


This scenario was investigated beyond the framework of mean field
theory by M\'{a}ska {\it et al.}\cite{Maska}. By assuming that (i) the
cuprate superconductors can be described by the $t-J$ model, with
the exchange interaction as the main pairing
mechanism~\cite{Ruckenstein,Baskaran}, and (ii) that the presence
of the dopant atoms induces a position-dependent shift of the
atomic levels in the CuO$_2$ plane,  these authors calculated the
effective superexchange interaction $J$ between copper ions in the
presence of dopants from a perturbation expansion of the one-band
Hubbard Hamiltonian up to second order. They  showed that
 the diagonal disorder in the plane
{\it always} leads to an enhancement of $J$;  accordingly, with
the assumption that pairing is due to superexchange, the
superconducting gap increases in the vicinity of the dopant atoms,
in agreement with Ref. \onlinecite{Nunner_05}.  If true, this
conclusion would provide an important, apparently robust way to
connect local atomic displacements with the increase of the
pairing there, using the results of STS. However, recent results
by Johnston {\it et al.}~\cite{Johnston} based on cluster model
calculations for the three-band Hubbard Hamiltonian ~\cite{Emery},
which account explicitly for the Cu-O hopping processes, showed
instead that electrostatic modifications due to the presence of
the oxygen dopant locally suppress $J$.  They showed, in
addition, that electronic coupling to local phonon modes was
strongly modified by the dopant,  and could enhance $J$.  This is
consistent with the fact that the gap inhomogeneities are strongly
(anti)correlated to a local bosonic mode frequency identified in
the tunnelling conductance by Lee et al.\cite{jhlee}, but still in
apparent contradiction to the result of M\'aska {\it et al.}

In view of the present controversy, we investigate here the
possibility that the M\'aska {\it et al.} result is an artifact of an
oversimplified model of the electronic structure of the CuO$_2$
plane, and analyze  the effect of a dopant impurity
 on  $J$ by performing a
perturbation expansion on the three-band Hubbard model.  As in
Ref. \onlinecite{Maska}, we assume initially that the primary
effect of the dopant
is the shift of the atomic energy levels in the CuO$_2$ plane, but
account for the shifts in O levels as well as Cu. Our calculations
show that the fifth-order contribution is as important as the
fourth-order one, in agreement with the results for the pure case
\cite{Eskes}. We find that the sign of these contributions is very
susceptible to the relative dopant-induced spatial variation of
the atomic levels; in contrast to the single-band case, it may be
either positive or negative. Finally, we show how the
discrepancies between Refs. \onlinecite{Maska} and
\onlinecite{Johnston} may be understood in terms of limiting
considerations.

\section{ii. model}
\begin{figure}[tb]
\begin{center}
\includegraphics[width=5 cm]{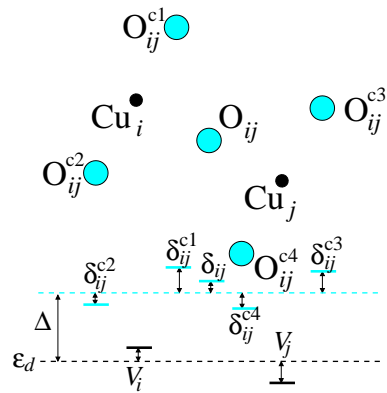}
\end{center}
\caption{(Color online) Energy level diagram for a Cu$_2$O$_5$ cluster illustrating the
notation used for the dopant-induced  shifts of Cu and O atomic energy levels.
 Each schematic energy level is located directly beneath its corresponding atom and accordingly
colored [black for Cu and cyan (gray) for O levels].} \label{En}
\end{figure}

Our starting model is the  three-band Hubbard Hamiltonian
$H_{\rm{Hub}}$
on the CuO$_2$ plane \cite{Emery}. The three bands
 arise from the hybridization of: the Cu 3$d_{x^2-y^2}$ orbital
and the two degenerate O 2$p$ orbitals, O 2$p_x$ and O 2$p_y$.
 In the  {hole} representation, $H_{\rm{Hub}}$ can be written as:
\begin{eqnarray}
H_{\rm{Hub}}&=&\sum_{i,\sigma}\left(\varepsilon_d+V_i\right)d^{\dagger}_{i,\sigma}d_{i,\sigma}
+\sum_{l, \sigma}\left(\varepsilon_d+\Delta+\delta_{l}\right)p^{\dagger}_{l,\sigma}p_{l,\sigma}\nonumber \\
&+&\sum_{<i,l>\sigma} t^{il}_{pd}\left(d^{\dagger}_{i,\sigma}p_{l,\sigma}+H.c\right)\nonumber\\
&+&\sum_{<l,n>\sigma} t^{ln}_{pp}\left(p^{\dagger}_{l,\sigma}p_{n,\sigma}+H.c\right)\nonumber \\
&+& U_d\sum_i d^{\dagger}_{i,\uparrow}d_{i,\uparrow}d^{\dagger}_{i,\downarrow}d_{i,\downarrow}
+U_p\sum_l p^{\dagger}_{l,\uparrow}p_{l,\uparrow}p^{\dagger}_{l,\downarrow}p_{l,\downarrow}.
\label{Ham}
\end{eqnarray}
 In Eq.~(\ref{Ham}), $d_{i,\sigma}^{\dagger}$
($d_{i,\sigma}$) creates (annihilates)
 a hole with spin $\sigma$ in the 3$d_{x^2-y^2}$ orbital of a Cu atom at
 site $i$. Correspondingly, $p_{l,\sigma}^{\dagger}$ ($p_{l,\sigma}$)
 creates (annihilates) a hole with spin $\sigma$ in one of the two O 2$p$
 orbitals at site $l$. $\varepsilon_d$ is the on-site energy
of the Cu $3d_{x^2-y^2}$ orbital, while $\Delta$ is the difference
 between the Cu $3d_{x^2-y^2}$  and the O $2p$ energies
 in the pure system. $t^{il}_{pd}$ and $t^{ln}_{pp}$ describe the nearest-neighbor
 Cu-O and  O-O hoppings, respectively. Only hoppings  within the CuO$_2$ plane
 are considered. The sign of $t^{il}_{pd}$ and $t^{ln}_{pp}$ depends on the relative phase of the overlapping
3$d_{x^2-y^2}$ and 2$p$ orbitals. $U_d$ ($U_p$) is the on-site Coulomb
 repulsion for a pair of holes on a Cu (O) atom. The presence
of a dopant
  shifts the atomic Cu and O energy levels in its neighborhood. We
denote the energy shift for a Cu at position $i$ as $V_i$
and for an O at position $l$ between Cu ions at positions
$i$ and $j$ as $\delta_l=\delta_{ij}$ (see Fig.~\ref{En}).
 Besides
 these shifts, the dopant
 is expected to cause local lattice distortions, which lead to
 the modification of the hopping integrals $t^{il}_{pd}$ and $t^{ln}_{pp}$.
In the present work we neglect this effect, as in Ref.
\onlinecite{Maska}, and
 concentrate on the effects due to the dopant-induced spatial
 variation of Cu and O atomic energy levels.

\section{iii. Perturbation expansion}
The fourth- and
fifth-order expressions for the superexchange interaction
($J^{(4)}$ and  $J^{(5)}$, respectively) in homogeneous cuprates
were derived by Eskes and Jefferson \cite{Eskes} using
Rayleigh-Schr\"odinger perturbation theory:
\begin{eqnarray}
 J&=&J^{(4)}+J^{(5)},\\
 J^{(4)}&=&\frac{4t^4_{pd}}{\Delta^2}\left\{\frac{1}{U_d}+\frac{2}{2\Delta+U_p}\right\},\label{J4}\\
 J^{(5)}&=&\frac{4t^4_{pd}}{\Delta^2}\left\{\frac{1}{U_d}\frac{8t_{pp}}{\Delta}
+\frac{2}{2\Delta+U_p}\frac{8t_{pp}}{\Delta}+\frac{4t_{pp}}{\Delta^2}\right\}\label{J5},
\end{eqnarray}
where we have set to zero the Coulomb repulsion $U_{pd}$ between a
hole on a Cu ion and a hole
 on the neighboring O ion, which we neglect in our calculations for the sake of simplicity.

We consider now the three-band Hubbard model for the  case with an
impurity [Eq.~(\ref{Ham})] in the regime where $V_i, \delta_l <
\Delta, U_d, U_p$ and apply Rayleigh-Schr\"odinger perturbation
theory. We treat the hopping terms in Eq. (\ref{Ham}) as a
perturbation $H_1$,
\begin{eqnarray}
 H_1&=&\sum_{<i,l>\sigma} t^{il}_{pd}\left(d^{\dagger}_{i,\sigma}p_{l,\sigma}+H.c\right)\nonumber \\
&+&\sum_{<l,n>\sigma} t^{ln}_{pp}\left(p^{\dagger}_{l,\sigma}p_{n,\sigma}+H.c\right).
\end{eqnarray}
The ground state of the unperturbed  Hamiltonian $H_{0}$
($H_{\rm{Hub}} = H_0 + H_1$)  corresponds in this case
to all Cu atoms occupied
by one hole each. This state is $2^N$-fold-degenerate due to the various possible electron spin distributions
\begin{equation}\displaystyle
 \left|\sigma_1\cdots\sigma_N\right>=\prod^N_{i=1}d^{\dagger}_{i,\sigma_i}\left|{\rm
vac}\right>,\; \label{GS}
\end{equation}
where $\sigma_1,\ldots,\sigma_N=\uparrow$ or $\downarrow$ and $i$ runs over Cu sites.

The effective Hamiltonian $H_{\rm{eff}}$ is calculated as a
perturbation expansion in powers of $H_1$ \cite{Lindgren,Takahashi,Mueller}.
For the set of states (\ref{GS}), we can ignore many terms of the
 perturbation series by making use of the fact that
the terms containing $PH_1P$, where the operator $P$ projects on the ground state manifold
 Eq.~(\ref{GS}), will all vanish since it is not possible to connect any two states out of
the ground state manifold [Eq.~(\ref{GS})] by a single hopping process. This observation leads to the
following expression for $H_{\rm{eff}}$:
\begin{eqnarray}
 H_{\rm{eff}}&=&E_0P+PH_1RH_1P+PH_1RH_1RH_1P \label{H_eff} \nonumber\\
  &+&PH_1RH_1RH_1RH_1P\nonumber \\
  &-&\frac{1}{2}PH_1R^2H_1PH_1RH_1P-\frac{1}{2}PH_1RH_1PH_1R^2H_1P\nonumber \\
  &+&PH_1RH_1RH_1RH_1RH_1P\nonumber \\
  &-&\frac{1}{2}PH_1RH_1R^2H_1PH_1RH_1P\nonumber\\
  &-&\frac{1}{2}PH_1RH_1PH_1R^2H_1RH_1P\nonumber\\
  &-&\frac{1}{2}PH_1R^2H_1RH_1PH_1RH_1P\nonumber\\
  &-&\frac{1}{2}PH_1RH_1PH_1RH_1R^2H_1P\nonumber\\
  &-&\frac{1}{2}PH_1R^2H_1PH_1RH_1RH_1P\nonumber\\
  &-&\frac{1}{2}PH_1RH_1RH_1PH_1R^2H_1P,
\end{eqnarray}
where
  $R=(1-P)/(E_0-H_0)$ so that, for a state
$\left|\phi\right>\cancel{\in} \left\{\left|\sigma_1\cdots\sigma_N\right>\right\}$,
\begin{equation}
 R\left|\phi\right>=\frac{1}{E_0-E_{\phi}}\left|\phi\right>.
\end{equation}
$E_0$ is the ground state energy of $H_0$ and $E_{\phi}=\left<\phi|H_0|\phi\right>$.

Among the terms in $H_{\rm{eff}}$, Eq.~(\ref{H_eff}),
we need only to consider those  terms that are of the form
\begin{equation}
 \sum_{<i,j>,\sigma}d^{\dagger}_{i,\sigma}d^{\dagger}_{j,\bar{\sigma}}d_{j,\sigma}d_{i,\bar{\sigma}},\label{important}
\end{equation}
with $\bar{\sigma}=-\sigma$, since the corresponding  prefactor  determines $J$.
 The terms of interest result from calculating
 the fourth-order term $PH_1RH_1RH_1RH_1P$ and the fifth-order
 term $PH_1RH_1RH_1RH_1RH_1P$ in Eq.~(\ref{H_eff}). All other terms will only add a constant energy term to the effective Hamiltonian.

It is convenient to use graphs for deriving expressions for $J^{(4)}$
 and $J^{(5)}$. One has to consider all possible fourth- and fifth-order
 hopping processes resulting in the exchange of spins between
two Cu atoms and sum up the corresponding $PH_1RH_1RH_1RH_1P$
and $PH_1RH_1RH_1RH_1RH_1P$ expressions.
\begin{figure}[tb]
\begin{center}
\subfigure {\includegraphics[width=3.25 cm]{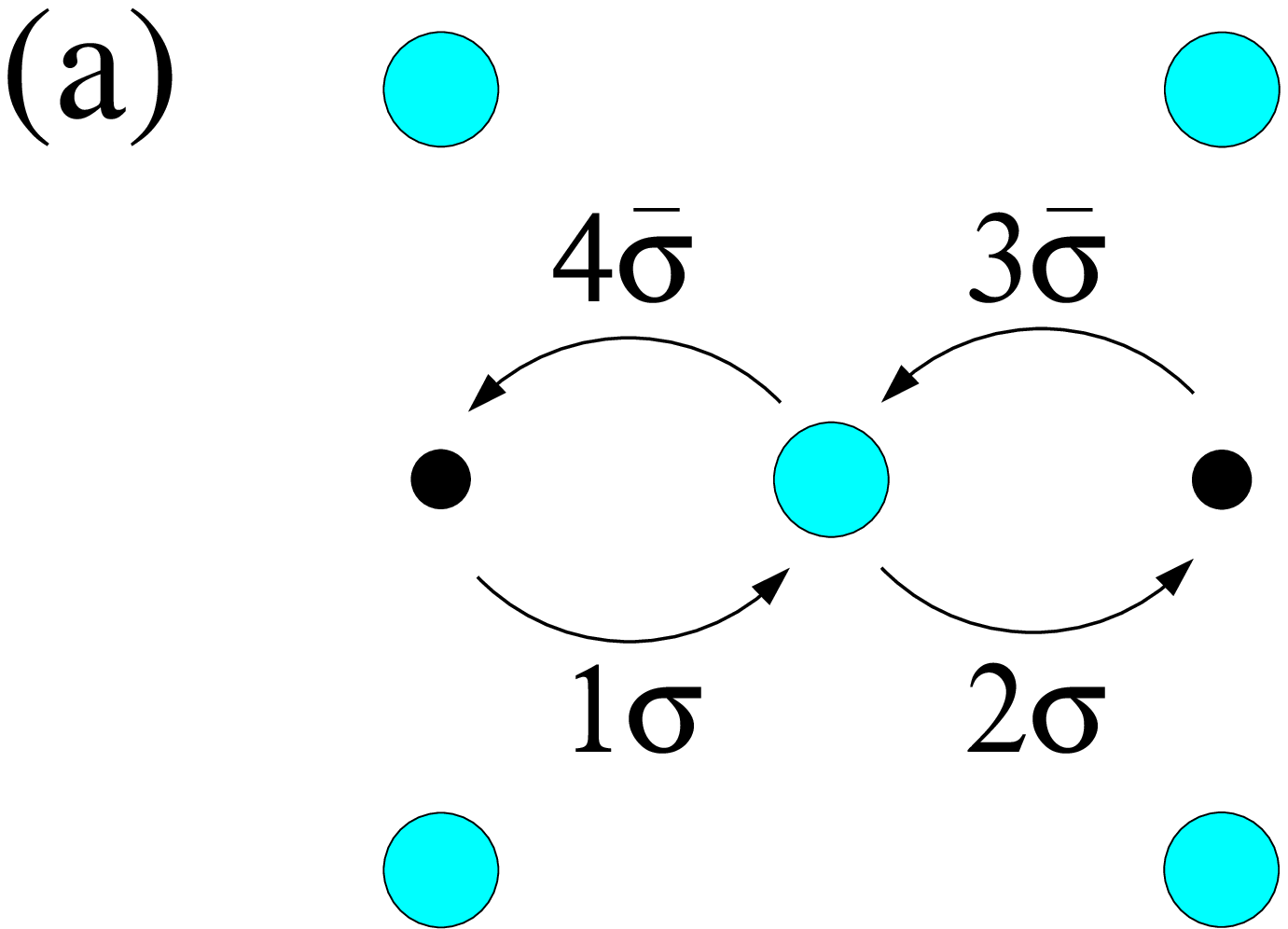}}\hspace{0.5cm}
\subfigure {\includegraphics[width=4.0 cm]{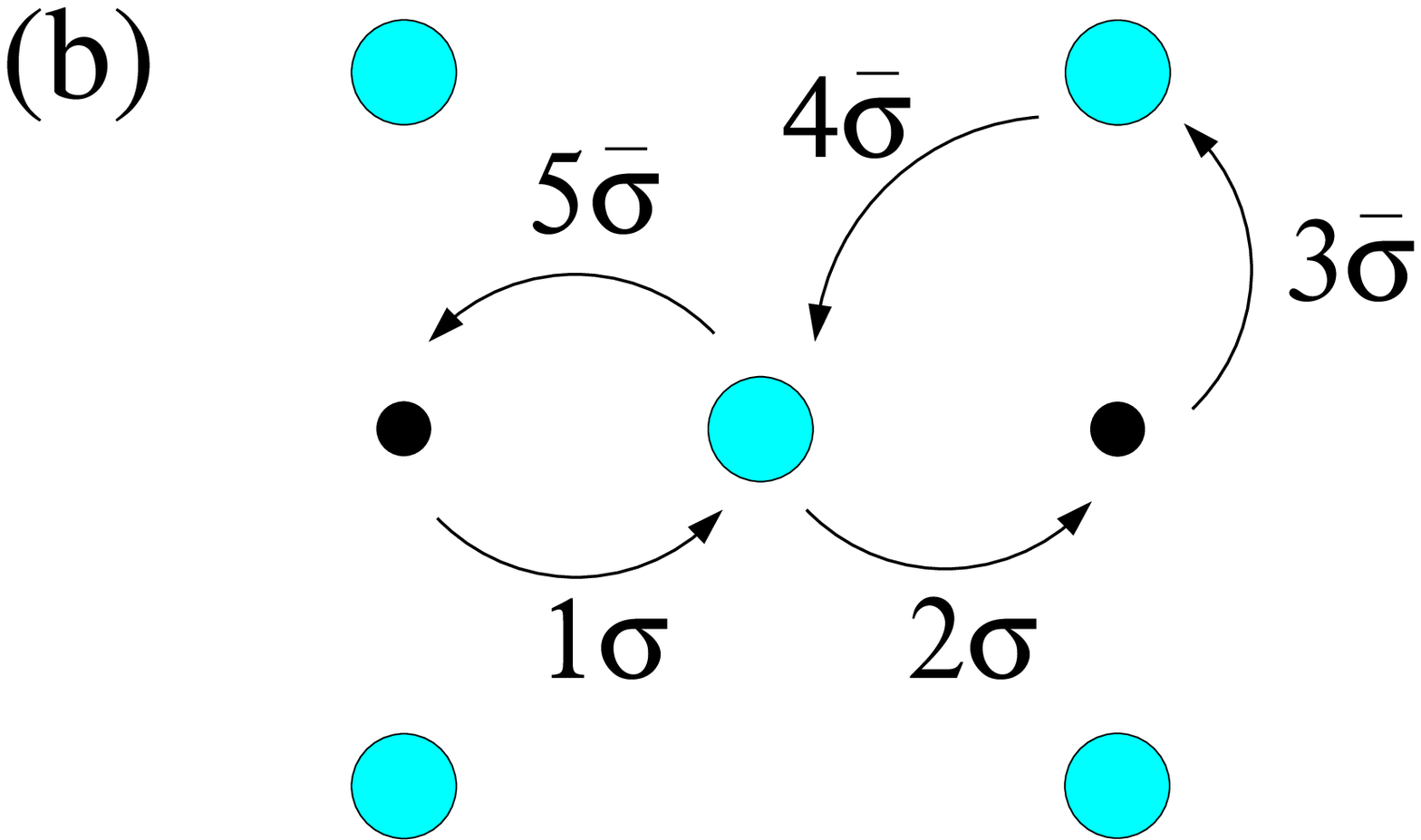}}
\end{center}
\caption{(Color online) Graphs describing the (a) fourth-  and the
(b) fifth-order  hole hopping processes that result in the
exchange
 of spins between two Cu atoms. Black [cyan (gray)] circles represent Cu [O] atoms. Arrows denote hopping processes,
 with the accompanying number indicating the order, in which the hoppings occur.  Symbols $\sigma$ or $\bar{\sigma}$
stand for the spin of the hole.}
\label{graph_4}
\end{figure}

There are in total 12 fourth-order graphs (each of the 6
topologically distinct graphs has two versions that differ by
flipped spins), one of which is shown in Fig. \ref{graph_4}(a).
In the pure case they reduce to only two terms in $J^{(4)}$ [Eq.
(\ref{J4})]. In the case of an impurity (dopant-induced spatial
variation of Cu and O levels), the terms in the sum for $J^{(4)}$
corresponding to  hoppings that start from the Cu ion at site $i$
will differ from those corresponding to hoppings that start from a
Cu ion at site $j$ due to the different dopant-induced shifts
$V_i$ and $V_j$. With the notation $t_{pd}=|t_{pd}^{il}|$,
$t_{pp}=|t_{pp}^{ln}|$, the local exchange $J^{(4)}_{ij}$ to
fourth-order is then
\begin{eqnarray}
 J^{(4)}_{ij}&=&\left(\frac{4t^4_{pd}}{\Delta^2}\frac{1}{U_d}+\eta^{(1)}_{ij}\right)\nonumber \\
&+&\left(\frac{4t^4_{pd}}{\Delta^2}\frac{2}{2\Delta+U_p}+\eta^{(2)}_{ij}\right),
\end{eqnarray}
with corrections
\begin{eqnarray}
\eta^{(1)}_{ij}&=&\frac{4t^4_{pd}}{\Delta^2}\nonumber\\
&\times&\frac{1}{U_d}\,\frac{a_0+a_1U_d+a_2U_d^2}{(\Delta-v_i)^2(\Delta-v_j)^2\left[U^2_d-(v_j-v_i)^2\right]},\\
&&\nonumber\\
a_0&=&(v_j-v_i)^2(\Delta-v_j)^2(\Delta-v_i)^2,\nonumber\\
a_1&=&\frac{1}{2}(v_j-v_i)^2(2\Delta-(v_i+v_j))\Delta^2,\nonumber\\
a_2&=&\frac{1}{2}(\Delta-v_j)^2(2\Delta-v_i)v_i
 +\frac{1}{2}(\Delta-v_i)^2(2\Delta-v_j)v_j.\nonumber\label{eta1}
\end{eqnarray}
and
\begin{eqnarray}
 \eta^{(2)}_{ij}&=&\frac{4t^4_{pd}}{\Delta^2}\,\frac{2}{2\Delta+U_p}\label{eta2}\\
&\times&
\frac{b_0+b_1U_p}{\left[(2\Delta-v_i-v_j)+U_p\right](\Delta-v_j)^2(\Delta-v_i)^2},\nonumber\\
&&\nonumber\\
b_0&=&
(\Delta-v_j)^2(\Delta^2+\Delta(\Delta-v_i)+(\Delta-v_i)^2)v_i\nonumber\\
&+&(\Delta-v_i)^2(\Delta^2+\Delta(\Delta-v_j)+(\Delta-v_j)^2)v_j\nonumber\\
&-&\frac{1}{2}\Delta^3(v_j-v_i)^2,\nonumber\\
&&\nonumber\\
b_1&=& \frac{1}{2}\left[(\Delta-v_j)v_i+(\Delta-v_i)v_j\right]\nonumber\\
&\times&
\left[\Delta\left(\Delta-\frac{v_i+v_j}{2}\right)+(\Delta-v_i)(\Delta-v_j)\right],\nonumber
\end{eqnarray}
where we have defined $v_i=V_i-\delta_{ij}$ and
$v_j=V_j-\delta_{ij}$.  It is easy to check that the
correction terms $\eta_{ij}^{(1)}$ and $\eta_{ij}^{(2)}$ vanish
when the impurity-induced potentials $v_i$ vanish.

\begin{figure}[tb]
\begin{center}
\subfigure {\includegraphics[width=6 cm]{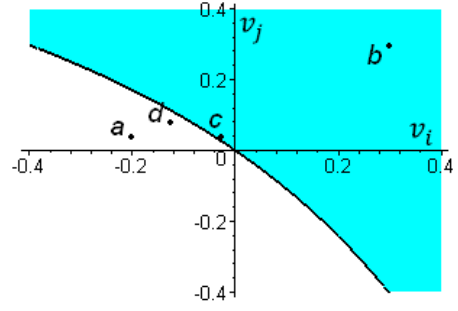}}
\end{center}
\caption{(Color online) sign($\eta_{ij}$) diagram for a typical set of the model parameters
[$U_d=8.8$ eV, $U_p=4.1$ eV and $\Delta=2.92$ eV (Ref.~\onlinecite{Johnston})]
 in the space of abscissa $v_i=V_i-\delta_{ij}$ and ordinate
$v_j=V_j-\delta_{ij}$. White and cyan (gray)  regions indicate
 negative and positive total fourth-order correction to $J$, respectively. Points (a)-(d) denote
the values of $v_i$ and $v_j$ used for generating diagrams (a)-(d) in Fig. \ref{5th}.}
\label{4th}
\end{figure}

 We  note that the sign of the total fourth-order correction due
to the presence of the dopant,
 $\eta_{ij}=\eta^{(1)}_{ij}+\eta^{(2)}_{ij}$, depends on the sign
and the magnitude of $v_i$ and $v_j$, i.e.,
 the actual energy separation between the dopant-shifted Cu and O levels and, in particular,
for $v_i, v_j<<\Delta$, $ \eta_{ij}$ is proportional to
$v_i+v_j$. In general,
 this result is shown
 diagrammatically in Fig. \ref{4th}  for a typical set of  model parameters in the Bi superconductors
[$U_d=8.8$ eV, $U_p=4.1$ eV and $\Delta=2.92$ eV (Ref.~\onlinecite{Johnston})]
. In the space of $v_i,v_j$, negative  and positive $\eta_{ij}$
contributions to $J$  are shown as white and gray (cyan) areas,
respectively.
  For the parameters considered in the cluster calculation
of Ref.~\onlinecite{Johnston}, the contribution of $\eta_{ij}$ is negative (black dot (d)
in Fig. \ref{4th}) and therefore the fourth-order correction suppresses $J$ in that case.

The number of graphs contributing to the fifth-order correction to the superexchange, $J^{(5)}$, is 120.
 They all involve hoppings to one of the four corner O ions [Fig. \ref{graph_4} (b)] so that $J^{(5)}$ will also depend on the corner O atoms energy level shifts $\delta_{ij}^{c1}$, $\delta_{ij}^{c2}$, $\delta_{ij}^{c3}$, and $\delta_{ij}^{c4}$ (Fig. \ref{En}). For example, the correction to the first term of $J^{(5)}$ in Eq.~(\ref{J5}) is
\begin{eqnarray}
 J^{(5)}_1&=&4t_{pd}^4t_{pp}\nonumber\\
&\times&\left(\frac{1}{\left(\Delta+\delta_{ij}-V_i\right)^2}\frac{1}{U_d+\{V_j-V_i\}}
\left\{\frac{1}{\Delta+\delta_{ij}^{c1}-V_i}\right.\right.\nonumber
\\
&+&\left.\frac{1}{\Delta+\delta_{ij}^{c2}-V_i}+\frac{1}{\Delta+\delta_{ij}^{c3}-V_i}+
\frac{1}{\Delta+\delta_{ij}^{c4}-V_i}\right\}\nonumber\\
&+&\frac{1}{\left(\Delta+\delta_{ij}-V_j\right)^2}\frac{1}{U_d-\{V_j-V_i\}}\left\{\frac{
1}{\Delta+\delta_{ij}^{c1}-V_j}\right.\nonumber\\
&+&\left.\left.\frac{1}{\Delta+\delta_{ij}^{c2}-V_j}+\frac{1}{\Delta+\delta_{ij}^{c3}
-V_j}+\frac{1}{\Delta+\delta_{ij}^{c4}-V_j}\right\}\right).\nonumber\\
\end{eqnarray}
The sign of this term depends on $v_i$, $v_j$, $v^{c1}_i=V_i-\delta^{c1}_{ij}$, $v^{c2}_i=
V_i-\delta^{c2}_{ij}$, $v^{c3}_j=V_j-\delta^{c3}_{ij}$, and $v^{c4}_j=V_j-\delta^{c4}_{ij}$
(alternatively, $v^{c2}_j=V_j-\delta^{c2}_{ij}$, etc. could be considered).
 These six parameters define the sign of the {\it total} fifth-order correction $\mu_{ij}$ as well.

\begin{figure}[tb]
\begin{center}
\subfigure [$v_i=-0.2$, $v_j=0.04$]{\includegraphics[width=4 cm]{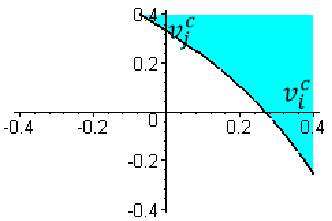}}
\subfigure [$v_i=0.3$, $v_j=0.3$]{\includegraphics[width=3.9 cm]{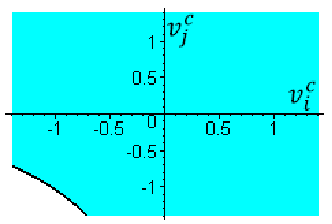}}
\subfigure [$v_i=-0.03$, $v_j=0.04$]{\includegraphics[width=4 cm]{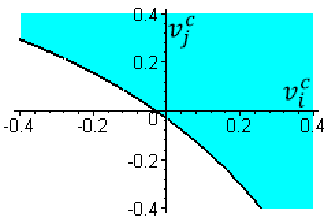}}
\subfigure [$v_i=-0.13$, $v_j=0.08$]{\includegraphics[width=4 cm]{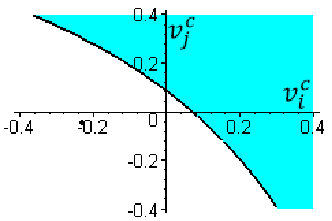}}
\end{center}
\caption{(Color online) sign($\eta_{ij}+\mu_{ij}$) diagrams in the
space of abscissa $v^c_i$ and ordinate $v^c_j$. White and cyan
(gray) correspond to negative and positive values of
$\eta_{ij}+\mu_{ij}$, respectively. The point $v^c_i$=-0.23 eV,
$v^c_j$=-0.05 eV in diagram (d) corresponds to the energy levels
distribution shown in Fig. 2 of Ref. \onlinecite{Johnston}.}

\label{5th}
\end{figure}

In order to quantify the effect
 that hopping to the corner O atoms has on the superexchange $J$, we
 consider the case  where $v^{c1}_i=v^{c2}_i=v^{c}_i$ and $v^{c3}_j=v^{c4}_j=v^{c}_j$ (such a symmetry is realized when the dopant atom is located on the line connecting two Cu atoms \cite{Johnston}). For given $v_i$ and $v_j$, it is then possible to draw a phase diagram of the sign of the total correction, $\mu_{ij}+\eta_{ij}$, in the space of $v^{c}_i$ and $v^{c}_j$. In Fig. \ref{5th}, we present, as an example, four such diagrams corresponding to different sets of $v_i$ and $v_j$ (four points in Fig. \ref{4th}). For calculating these diagrams we chose
 $t_{pd}$=1.2 eV and $t_{pp}$=0.5 eV as also considered in the cluster calculations
Ref.~\onlinecite{Johnston}. The local Cu and O site energies calculated by Johnston {\it et al.}
 correspond to the choice of $v_i$=-0.13 and $v_j$=0.08 eV in Fig. \ref{5th} (d).
 It can be concluded from examining the diagrams in Fig. \ref{5th},
that the parameters $v^{c1}_i$, $v^{c2}_i$, $v^{c3}_j$, and $v^{c4}_j$ have to be
slightly larger  than $v_i$ and $v_j$ to induce
 the change of sign of the correction to $J$ as compared to the fourth-order  result.

\section{iv. Analysis and Discussion}
We are now concerned
with trying to answer the following questions. Is the sign of the correction to $J$ caused by the
impurity uniformly positive, as occurred in the one-band calculation of Ref.
\onlinecite{Maska}?
 For physically reasonable assumptions
regarding the magnitude and spatial dependence of the impurity
potential for a dopant sitting several \AA~from the CuO$_2$ plane,
can the modulation of $J$ be significant at all?

Regarding the first point, there is a simple argument that
explains why the superexchange corrections due to doping derived
from the three-band model can assume both positive and negative
values while from the one-band model one  finds
 that $J$  is always enhanced. Let us consider what happens between neighboring
 Cu and O ions when their  energy levels shift due
 to a dopant by $V$ and $\delta$, respectively.
 The local separation between their energy levels,
which we denoted as $\Delta$ for the homogeneous case,
 varies as $\Delta_{\rm{loc}}=\Delta-(V-\delta)$. For $(V-\delta)>0$,
 $\Delta_{\rm{loc}}$ decreases compared with $\Delta$ and vice versa.
 Expressing $J$ in terms of $\Delta_{\rm{loc}}$ instead of $\Delta$,
 one sees that for $V_j-V_i\ll\Delta_{\rm{loc}}$ [in this limit $\Delta$ in Eq.~(\ref{J4})
 and Eq.~(\ref{J5}] can be replaced by $\Delta_{\rm{loc}}$) the variation
 of local Cu and O levels separation $\Delta_{\rm{loc}}$ defines the change
 of $J$: since $\Delta_{\rm{loc}}$ is in the denominator, $(V-\delta)>0$ leads
to the enhancement of $J$ and $(V-\delta)<0$ leads to the
 suppression of $J$. In Fig. \ref{4th}, the condition $V_j-V_i\ll\Delta_{\rm{loc}}$
 is fulfilled  in the vicinity of the $v_i=v_j$ line (on the line, $V_j-V_i=0$)
 and indeed $J$ is increased  in the first quarter and reduced in the third
 quarter of the diagram.
 In the second and fourth quarters the relative variation
 of levels of Cu atoms, $V_i-V_j$, becomes equally important. The one-band model excludes completely
the O atoms, thus ignoring one of the two microscopic factors
(change in the Cu-O energy levels separation and the relative
shift of the energy levels of two interacting Cu ions)
 that govern the variation of the local superexchange coupling $J$.

We would like to note that,
 as shown by Eskes and Jefferson \cite{Eskes} for the homogeneous case,
 even the fifth-order perturbation expansion for $J$ is insufficient for quantitative estimates
of $J$  and gives overestimated values compared with the
experimental (and cluster-model calculated) values of $J$. Such
trends, naturally, are also to be expected in the disordered case.
We calculate the value of the forth- and fifth-order superexchange
coupling corrections with the same model parameters as used in the
cluster-model calculations \cite{Johnston} [Fig. \ref{4th} and
\ref{5th} (d)] and find that 
 the sign and the order of magnitude of the correction
within our calculation, yielding a suppression of $J$ by ${\cal
O}(5$\%), are in good agreement with cluster
calculations\cite{Johnston} when no modulation in the hopping
integrals is considered, as it is our case here. Consideration of
other sets of model parameters\cite{Kent08} lead to the same
relative correction values.

\section{v. conclusions}
The question of the impact of a dopant atom on the local
electronic properties in the CuO$_2$ plane of the cuprates has
been highlighted by STM measurements, indicating that dopants
correlate with regions of large gap \cite{KMcElroy_05}, and the
theoretical proposal that the dopant itself is  enhancing the
pairing interaction \cite{Nunner_05}.  The appealing argument of
Ma\'{s}ka {\it et al.} \cite{Maska}, based on a single-band
analysis, that the perturbation provided by the dopant necessarily
enhances the superexchange locally and may therefore enhance
pairing, has been shown to be a special result restricted to
one-band systems. Within a three-band Hubbard model appropriate to
the CuO$_2$ plane, we have in this work performed a perturbative
calculation to fifth order in the hoppings $t_{pd}$ and $t_{pp}$,
and shown that the sign of the correction to $J$ can be positive
or negative depending on the potentials on nearby sites induced by
the dopant impurity. The typical modulation is of order
$d\Delta/\Delta$ times the exchange for the homogeneous system,
where $d \Delta$ is a typical dopant-dependent modulation of  the
local charge transfer energy  between Cu and O, and $\Delta$ is
the homogeneous value of this difference. Using values of these
shifts obtained from cluster calculations \cite{Johnston}, we find
that a typical modulation due to an O dopant in the Bi-2212 system
imaged by STM  is of order 5\% of the homogeneous value.

  It is possible
that a more accurate microscopic calculation, accounting for the
modulations of the hoppings and the apical oxygen degrees of
freedom neglected here
 may produce a reliable description
of this modulation.   Until then, we have shown that the size and
sign  of this modulation is not universal but depends on details
of the impurity and local electronic structure.

\section{Acknowledgements}

We acknowledge useful discussions with T.
P. Devereaux.  KF and RV thank the German Science Foundation
(DFG) for financial support through the  SFB/TRR49 program.  PJH
thanks DOE DE-FG02-05ER46236 for partial support.





\end{document}